# Gower's Similarity Coefficients with Automatic Weight Selection

*Marcello D'Orazio*
*Italian National Institute of Statistics - Istat, Rome, Italy*
*Via Cesare Balbo, 16 – 00184 Rome (RM) - ITALY*
*marcello.dorazio@istat.it*

**Abstract**

Nearest-neighbor methods have become popular in statistics and play a key role in statistical learning. Important decisions in nearest-neighbor methods concern the variables to use (when many potential candidates exist) and how to measure the dissimilarity between units. The first decision depends on the scope of the application while second depends mainly on the type of variables. Unfortunately, relatively few options permit to handle mixed-type variables, a situation frequently encountered in practical applications. The most popular dissimilarity for mixed-type variables is derived as the complement to one of the Gower's similarity coefficient. It is appealing because ranges between 0 and 1, being an average of the scaled dissimilarities calculated variable by variable, handles missing values and allows for a user-defined weighting scheme when averaging dissimilarities. The discussion on the weighting schemes is sometimes misleading since it often ignores that the unweighted "standard" setting hides an unbalanced contribution of the single variables to the overall dissimilarity. We address this drawback following the recent idea of introducing a weighting scheme that minimizes the differences in the correlation between each contributing dissimilarity and the resulting weighted Gower's dissimilarity. In particular, this note proposes different approaches for measuring the correlation depending on the type of variables. The performances of the proposed approaches are evaluated in simulation studies related to classification and imputation of missing values.

**Keywords**: donor-based methods; hotdeck imputation; similarity/dissimilarity.

# 1. Introduction

Dissimilarities or distances between observations have been extensively used in clustering applications to separate units into homogenous groups (see e.g. Kaufman and Rousseeuw, 1990). Recently, dissimilarity/distance-based methods have become popular in the area of *supervised statistical learning*, in both classification and regression problems (clustering can be seen as *unsupervised learning*). As noted by Hastie *et al.* (2009, p. 459), the popularity originates from their simplicity if compared to fitting complex statistical models. In particular they are very effective in classification problems and in regression applications, but when the objective is that of getting predictions and not understanding the nature of the relationship between the predictors and the response variables. Dissimilarity-based methods are very popular also in imputing the missing values (cf. Andridge and Little, 2010) or in data integration problems (*deterministic record linkage* and *data fusion*; cf. Coli *et al*, 2016; D'Orazio *et al.*, 2006). In imputation problems a missing value is replaced with the value observed on the closest responding unit (*nearest neighbor donor hotdeck*); this approach avoids fitting models and replacing the missing values with "artificial" model predictions.

It is worth noting that often the terms "distance" and "dissimilarity" are used interchangeably although a slight difference exists; a distance function must satisfy: non-negativity, symmetry and triangle inequality properties (see e.g. Kaufman and Rousseeuw, 2005). Dissimilarity is a wider concept because it is a function that may not follow the triangle inequality, in addition it is often bounded by unity. In this note we focus on measuring dissimilarity between observations in presence of mixed type variables; this is quite common when working with data collected in a survey on households or single individuals.

Although dissimilarity-based methods appear simple, their application and the corresponding results depend on two key choices: (i) which of the available variables should be used to calculate the dissimilarity (or the distance) and, (ii) how to measure dissimilarity (or distance). The decisions are strictly related; the first relates to the scope of application; for instance, in imputation of missing values the variables to select are the best predictors of the variable to impute but also capable to explain the nonresponse "mechanism" (see e.g. Andridge and Little, 2010). Often, the parsimony is the main guiding principle because the capacity of discriminating between close and far observations diminishes as the number of variables increases, i.e. everything starts being close to everything; this effect is known as *concentration effect* (see e.g. Zimek et al, 2012), and more in general falls under the *curse of dimensionality* (cf. Marimont and Shapiro, 1979). The selection of the variables is beyond the scope of the application and is not tackled here.

How to measure the dissimilarity (or distance) mainly depends on the type of selected variables according to the measurement scale. A *categorical nominal variable* can assume a finite number of categories that do not have a natural ordering; when it presents just two categories (yes/no, presence/absence, smoker/non-smoker, etc.; often represented by '0' and '1' where commonly '0'=absence and '1'=presence) it is also identified as *binary* variable.

A *categorical ordinal variable* shows a finite number of categories with a natural ordering, but the difference between categories is not meaningful. Variables measured on an *interval-scale* show quantitative measurement (integer or real numbers) and the difference between values is meaningful, but not the ratio (multiplication). Finally, a variable measured on a *ratio-scale* is an interval-scaled variable that admits the zero and where the ratio is meaningful. Often a categorical variable is identified as *qualitative* while an interval or ratio-scale variable is said *quantitative* or *continuous*.

When the selected variables are all categorical or all non-categorical, many alternative ways of measuring dissimilarity (or distance) are available (see e.g. Chapter 1 in Kaufman and Rousseeuw, 2005; or Chapter 3 in Podani, 2000). Unfortunately, relatively few options can handle mixed-type variables; the most popular is the complement to one of the *Gower's similarity coefficient* (Gower, 1971); it is an average of the dissimilarities calculated variable by variable, whereas each dissimilarity is bounded by 1 (maximum dissimilarity). The Gower's similarity is appealing because handles missing values and allows a user-defined weighting scheme when averaging dissimilarities over the single variables. Unfortunately, the discussion on the weighting schemes often ignores that in absence of weights (unweighted "standard" version, that corresponds to a uniform weighting scheme) the categorical variables have a higher impact on the final similarity if compared to numeric ones.

Other approaches to measure distance or dissimilarity in presence of mixed-type variables have been introduced in clustering applications; most of them suggest applying a linear combination of the dissimilarities calculated on subgroups of variables of the same type. Some of them can be viewed as extensions of the *K-means* (see e.g. van de Velden *et al.*, 2018; Ahmad and Khan, 2019). An issue that limits the applicability of these proposals beyond clustering problems is the fact that the weights assigned to the different components of the overall dissimilarity are typically optimized taking into the characteristics of the clustering procedure and often they are not kept constant but updated in the various steps. Finally, many of the proposed approaches do not permit to handle missing values, contrarily to the Gower's dissimilarity.

It is worth noting that when calculating dissimilarity with mixed-type variables some practitioners bypass the problem by replacing each categorical variable with the corresponding dummies and treat them as quantitative variables (i.e. binary variables are treated as continuous) and consequently apply functions designed to handle continuous variables; it is a common mistake that has no foundations and practically returns not meaningful dissimilarity measures. This practice is very common with binary variables whose categories are coded with "0" (absence) and "1" (presence) although may have undesired consequences (see. e.g. Section 2.4 in Kaufman and Rousseeuw, 2005) [1]. In alternative, sometimes the interval and ratio-scaled variables are categorized to apply distance functions tailored to handle categorical variables; this choice has the advantage of returning meaningful dissimilarities but the categorization implies a loss of information and the decision about the number of categories is subjective and not straightforward.

When using dissimilarity-based methods in imputation or data integration problems, an alternative solution to handle mixed-type variables often consists in calculating "conditional dissimilarities". In practice, initial groups of units (*donors' pools*) are created by considering units sharing the same value(s) for the chosen categorical variable(s) then the dissimilarity between units in the same pool is calculated using the remaining quantitative variables. The dissimilarity between units belonging to different groups is not calculated (see e.g. Andridge and Little, 2010) and practically corresponds to set it to the maximum achievable value. This way of working assigns explicitly a higher importance to categorical variables.

---

[1] This problem is also be tackled in Section 2 when discussing the reasons for discriminating between *binary symmetric* and *binary asymmetric* variables when calculating the Gower's similarity.

This work focuses on the Gower's dissimilarity because it has the advantage of being applicable for solving different problems (clustering, imputation, etc.). We address the drawback of the unbalanced contribution of the different types of variables to the overall unweighted Gower's dissimilarity by applying a non-uniform weighting mechanism that, following the suggestion of de Bello et al (2021), achieves a balance in terms of correlation between the dissimilarity calculated on each single variable and the final weighted Gower dissimilarity. In particular, this note suggests some improvements in measuring the correlation between a continuous and a binary variable, exploring also the possibility of replacing linear correlation with rank's correlation.

The remaining of the article is organized as follows. The Section 2 illustrates the main characteristics of the Gower's similarity. Section 3 shows how the different types of variables contribute to the overall unweighted Gower's dissimilarity and introduces some approaches to automatically select a weighting scheme in order to compensate this drawback. The behavior of the suggested approaches to choose the weights is investigated in a series of simulation studies whose results are presented in the Section 4. Finally, the Section 5 summarizes the main findings.

## 2. The Gower's dissimilarity

The Gower's (1971) proposal is the most popular way of measuring the similarity/dissimilarity between observations in the presence of mixed-type variables. The *Gower's dissimilarity* can be defined as the complement to one of the Gower's *similarity coefficient*:

$$d_{G,ij} = 1 - s_{G,ij} = \frac{\sum_{t=1}^{p} \delta_{ijt} \, d_{ijt}}{\sum_{t=1}^{p} \delta_{ijt}}$$

It is a dissimilarity measure (see e.g. Kaufman and Rousseeuw, 2005, p. 35) between unit $i$ ($i = 1,2, \ldots, n$) and unit $j$ ($j = 1,2, \ldots, n$) where $d_{ijt} = 1 - s_{ijt}$ is the dissimilarity calculated on the *t*th variable ($d_{iit} = 0$); $s_{ijt}$ is the similarity between $i$ and $j$ with respect to the *t*th variable and its value depends on the type of the variable itself ($s_{iit} = 1$). Table 1 shows how to calculate $s_{ijt}$ (and consequently of $d_{ijt} = 1 - s_{ijt}$) in correspondence of the different types of variables, being $x_{it}$ the value observed for the *t*th variable ($t = 1,2, \ldots, p$) on the *i*th unit.

Table 1 – Calculation of the Gower's similarity by type of variable

| Type of variable | $s_{ijt}$ | $\delta_{ijt}$ | Note |
|---|---|---|---|
| Binary symmetric | 1 if $x_{it} = x_{jt}$<br>0 if $x_{it} \neq x_{jt}$<br>0 if $x_{it}$ or $x_{jt}$ or both are missing | 1 if both the variables are nonmissing<br>0 if $x_{it}$ or $x_{jt}$ or both are missing | $s_{ijt}$ corresponds to the *simple matching coefficient* |
| Binary asymmetric | 1 if $x_{it} = x_{jt} = 1$<br>0 otherwise<br>0 if $x_{it}$ or $x_{jt}$ or both are missing | 1 if both the variables are nonmissing<br>0 if $x_{it} = x_{jt} = 0$<br>0 if $x_{it}$ or $x_{jt}$ or both are missing | $s_{ijt}$ corresponds to the *Jaccard index* |
| Categorical nominal (more than two categories) | 1 if $x_{it} = x_{jt}$<br>0 if $x_{it} \neq x_{jt}$<br>0 if $x_{it}$ or $x_{jt}$ or both are missing | 1 if both the variables are nonmissing<br>0 if $x_{it}$ or $x_{jt}$ or both are missing | $s_{ijt}$ is the simple matching on the untransformed variable or to the *Dice* (*Czekanovsky-Sorerensen*) *measure* applied to the dummies obtained by transforming the original variable |
| Measured on an interval or ratio scale | $1 - \|x_{it} - x_{jt}\|/R_t$<br>0 if $x_{it}$ or $x_{jt}$ or both are missing | 1 if both the variables are nonmissing<br>0 if $x_{it}$ or $x_{jt}$ or both are missing | $R_t = \max(x_t) - \min(x_t)$ is the range of the *k*th variable<br>$1 - s_{ijt}$ is *the Manhattan* or *city-block distance* scaled by the range |

As shown in Table 1, the Gower's coefficient treats in a different manner *binary symmetric* and *binary asymmetric* variables. This distinction reflects the different origin of the variable; a binary symmetric is a variable that can assume only two categories, e.g. smoker/nonsmoker. On the contrary, a binary asymmetric variable is the result of a "rough" observation or a transformation of a categorical nominal variables admitting more than two categories; consider for instance to observe the educational level just by asking whether an individual has achieved or not a university degree; when both the units have a university degree (both "1"=presence) their similarity is 1; on the contrary, when they do not share the university degree then they cannot be considered similar as they can have achieved different education levels (primary vs. secondary school etc.). Unfortunately, this distinction is often not considered in applications or in the implementation of the Gower's similarity in statistical software packages. It is worth noting that the Gower's seminal proposal (1971) when mentioning "dichotomous characters" refers to binary asymmetric variables, but later states that for "qualitative characters with two levels (i.e. alternatives)" the simple matching coefficient can be applied.

The Gower's proposal does not provide guidance for categorical ordered variables. Kaufman and Rousseeuw (2005) suggest to replace the categories with the corresponding position index in their natural ordering ($o_{it}$; $1 \leq o_{it} \leq C$) and then derive a new variable:

$$z_{it} = \frac{o_{it} - 1}{\max(o_t) - 1}$$

that, when calculating the similarity is treated as a variable measured on a ratio scale.

Podani (1999) replaces the ordered categories with the corresponding ranks ($r_{it}$) and then suggests to calculate the similarity as follows:

$$s_{ijt} = 1 - \frac{|r_{it} - r_{jt}|}{\max(r_t) - \min(r_t)}$$

This latter expression requires a correction to account for tied ranks (see Podani, 1999).

In practice, the Gower's dissimilarity ($d_{G,ij} = 1 - s_{G,ij}$) can be viewed as an average of the dissimilarities ($d_{ijt} = 1 - s_{ijt}$) measured on the available variables, where the dissimilarity calculated on each single variable is bounded by unity (maximum dissimilarity); as a consequence, the averaging provides an overall dissimilarity with values between 0 and 1 ($0 \leq d_{G,ij} \leq 1$).

The Gower's dissimilarity allows for missing values, which do not contribute to the calculation of the overall dissimilarity (dummy variable $\delta_{ijt}$). Obviously, if a unit presents missing values for all the $p$ variables then the dissimilarity with any other unit, fully or partially observed, would be undefined; for this reason, a unit with all the values missing should be discarded in advance.

A generalization of the Gower's dissimilarity consists in using a weighted average:

$$d_{wG,ij} = \frac{\sum_{t=1}^{p} \delta_{ijt} \, d_{ijt} \, w_t}{\sum_{t=1}^{p} \delta_{ijt} \, w_t}$$

that assigns a different weight, $w_t$, to each of the variables (the unweighted version corresponds to setting $w_t = 1$ for all the variables); Gower (1971) notes that the decision on a rational set of weights is difficult. Generally speaking, the discussion about potential alternative set of weights is often flawed by the thinking that the unweighted Gower's coefficient, assigning an equal weight to each variable ($w_t = 1$), corresponds to a balanced contribution of the various variables to the overall dissimilarity. In practice, this is not completely true and, in addition, the presence of outliers in numerical variables influences their contribution to the final overall dissimilarity; this is shown in the next Section by means of simple examples.

## 3. The contribution of different types of variables to the Gower's dissimilarity

The Gower's dissimilarity introduces a normalization of the single dissimilarities to have them ranging between 0 and 1; this operation is quite straightforward but can have a number of undesired consequences. Let's focus attention on the simple case of two variables: a categorical nominal variable and a quantitative one; with a categorical variable $d_{ijt} = 0$ for all the couples of observations sharing the same category ($x_{it} = x_{jt}$); this situation occurs less frequently for the quantitative variable where $d_{ijt} = 0$ only when $|x_{it} - x_{jt}| = R_t$. In other words, the standard unweighted Gower's dissimilarity shows an unbalanced contribution of quantitative variables compared to categorical ones and the problem is not trivial since the final dissimilarity depends: (i) on the number of categorical variables compared to continuous ones; (ii) on the number of categories; and, (iii) on the distribution of each categorical variable (see e.g. Foss *et al.*, 2016; Pavoine *et al.*, 2009).

To better illustrate the problem let's see how the Gower's dissimilarity between two individuals changes according to different combinations of age and smoking behavior (smoker/nonsmoker). Two individuals sharing the same smoking behavior but with a huge dissimilarity on age (15 vs 78; with $R_{age} = 85$) return $d_G = 0.3706 \ (= 0 + |15 - 78|/85)$ and are closer than two units having the same age but a different smoking behavior, as in this latter circumstance $d_G = 0.5 \ (= 1 + |24 - 24|/85)$; in practice, in identifying the closer units the Gower's dissimilarity tends to favor the observations sharing the same smoking behavior rather than the same age. Note that $d_G = 0.5$ also when the individuals have the same smoker behavior but their dissimilarity in terms of age is maximum (15 vs 100, $d_G = 0 + |15 - 100|/85 = 0.50$). In practice, $d_{age} = 1$ is achieved only when comparing the two extreme values in the opposite tails of the distribution (as observed on the available sample), where the age equal to 100 can be considered quite rare. This reveals an additional problem of the range-scaled Manhattan dissimilarity: the presence of outliers in the distribution of the continuous variable may affect directly the estimation of the range ($R_t$) and, as a consequence, the values of $d_{ijt}$ tend to be become smaller than one and $d_{ijt} = 1$ would be observed only rarely. The problem of outliers in numerical variables is not tackled here but some simple corrections can help in solving the problem, as shown in D'Orazio (2021).

The unbalanced contribution of the different types of variables is further exacerbated when the number of categorical variables increases. For instance, if the marital status is considered, in addition to smoking behavior and age, then individuals sharing the same smoking behavior and marital status have a dissimilarity smaller than that of two individuals with a close age but a different smoking behavior or marital status. In practice, the Gower dissimilarity identifies as closer the units sharing the same categories of the categorical variable, caring less on their dissimilarity on the ratio-scaled variable. This behavior can only be compensated by introducing a series of modifications by modifying the way of measuring the dissimilarity for the variables measured on an interval or ratio scale, as proposed by D'Orazio (2021), or by adopting a non-uniform weighting scheme so to assign a higher weight $w_t$ to the interval or ratio-scaled variables.

Gower (1971) suggests that "weights may be regarded as a function of the result of the values of a character being compared"; "differences in a character may be considered more important than agreement, or agreement between rare character states might be given more weight than agreement between common states"; this reasoning applies quite well to categorical variables where weights can be obtained as a function of the Shannon information (cf. Gower, 1971).

Use of unequal weights is common in clustering applications to assign higher weights to variables that can be used to separate the data into clusters whilst variables that poorly contribute to discriminate receive lower weights. In this article the objective is slightly different, being that of balancing the contribution of different types of variables to the overall dissimilarity. Pavoine et al (2009) illustrate the problem of unbalanced contribution of different types of variables by looking at the linear correlation between the dissimilarity calculated on each single variable ($d_{ijt}$) and the overall dissimilarity ($d_{wG,ij}$). Following this approach, de Bello et al (2021) suggest to determine automatically the weights ($w_t$) by minimizing the differences in the linear correlation between the dissimilarity calculated on each single variable ($d_{ijt}$) and the resulting weighted Gower dissimilarity ($d_{wG,ij}$). The problem can be solved analytically or by searching numerically the (approximate) solution through a *Genetic Algorithm* (belongs to the larger family of *evolutionary algorithms*; see e.g. Sivanandam and Deepa, 2007). The analytical solution can be achieved only in absence of missing values and when the set of variables does not include a binary asymmetric variable, i.e. when $\delta_{ijt} \neq 0$ and consequently the expression for calculating $d_{wG,ij}$ can be simplified in:

$$d_{wG,ij} = \frac{\sum_{t=1}^{p} d_{ijt} w_t}{\sum_{t=1}^{p} w_t} = \sum_{t=1}^{p} d_{ijt} w'_t$$

where the new weights sum up to 1 ($\sum_{t=1}^{p} w'_t = 1$). In this setting the analytical solution is achieved by solving a system of linear equations, under the additional constraint that there aren't couples of variables showing a linear correlation equal to one. Unfortunately, the analytical solution does not guarantee non-negative weights ($w'_t \geq 0$), this drawback can be avoided by applying the numerical approach that has the additional advantage of admitting missing values; the price to pay is a nonnegligible computational effort needed to reach the optimal solution (that practically minimizes the standard deviation of the various correlation coefficients between each single variable and the resulting weighted Gower dissimilarity); this effort increases with increasing number of variables, $p$, involved in the dissimilarity computation as well as the number of couples of units.

The idea of obtaining automatically a system of weights that balances the contribution of the different variables to the overall Gower's dissimilarity in terms of linear correlation is very appealing but has some drawbacks. First it is difficult to assume existence of monotonic linear relationship between the overall weighted dissimilarity and the dissimilarity calculated on each singe variable, in particular when the variable is categorical nominal. In this latter case the Gower's approach returns a dissimilarity assuming only a value equal to 0 or to 1, and its Pearson's product-moment correlation with the overall Gower's dissimilarity corresponds to the *point biserial correlation*:

$$r(d_{wG}, d_t) = \frac{\left(\bar{d}_{wG}^{(1)} - \bar{d}_{wG}^{(0)}\right)}{S_{d_{wG}}} \sqrt{\frac{m}{m-1} \bar{d}_t^{(1)} \left(1 - \bar{d}_t^{(1)}\right)}$$

In this latter expression $\bar{d}_{wG}^{(1)}$ is the average of the weighted Gower dissimilarities for the couples of units where $d_{ijt} = 1$, being $d_{ijt}$ the dissimilarity calculated on the *t*th categorical variable; similarly $\bar{d}_{wG}^{(0)}$ is the average corresponding to $d_{ijt} = 0$; $S_{d_{wG}}$ is the standard deviation (sample estimate) of the final weighted Gower's dissimilarity; finally, $\bar{d}_t^{(1)}$ is the proportion of the *m* couples where $d_{ijt} = 1$. Unfortunately, literature warns that the range of $r(d_{wG}, d_t)$ could be smaller than the usual $[-1,1]$ depending on the distribution of $d_{wG}$ (quantitative variable) and the proportion of 1s ($\bar{d}_t^{(1)}$) in the dichotomous variable $d_t$ (see Cheng and Liu, 2016); for instance, with a Gaussian distributed variable ($d_{wG}$) and a proportion of ones equal to 0.5 ($\bar{d}_t^{(1)} = 0.5$) it can be shown that the point biserial correlation ranges is [-0.798, 0.798]. In addition, $r(d_{wG}, d_t)$ is sensitive to departures from the assumption of equal variability of $d_{wG}$ in the two groups created according to the outcomes of $d_{ijt}$, that is implicitly considered when applying $r(d_{wG}, d_t)$. To partially compensate some of these drawbacks we suggest to adopt the modified estimator of biserial correlation proposed by Brogden (see Kraemer, 2006):

$$r_{mb}(d_{wG}, d_t) = \frac{\left(\bar{d}_{wG}^{(1)} - \bar{d}_{wG}^{(0)}\right)}{D_h}$$

that has the advantage of assuming solely a monotonic relationship between $d_{wG}$ and the dissimilarity $d_t$ calculated on a categorical nominal variable. In this latter expression

$$D_h = \frac{1}{h}\sum_{k=1}^{h} d_{wG(k)} - \frac{1}{m-h}\sum_{k=h+1}^{m} d_{wG(k)}$$

and is calculated starting from the values of $d_{wG}$ sorted decreasingly:
$$d_{wG(1)} \geq d_{wG(2)} \geq d_{wG(3)} \geq \cdots \geq d_{wG(m)};$$
*h* is an integer between 1 and *m*, the number of couple of units involved in dissimilarity computation, and is obtained as $h = \left[m \times \max\left(\bar{d}_{wG}^{(0)}, \bar{d}_{wG}^{(1)}\right)\right]$ (where [·] stands for integer part).

The assumption of linear correlation between dissimilarities can be avoided if rank correlation is considered; in fact, ranks permit to account for any monotonic relationship. In other words, another approach for setting for weights can be obtained by measuring the correlation between $d_{wG}$ and $d_t$: (i) by means of the Spearman's rank correlation coefficient when $d_t$ is continuous, while (ii) the *rank biserial correlation coefficient*:

$$r_{rb} = 2\frac{\bar{R}_1 - \bar{R}_0}{m}$$

should be applied when $d_t$ is dichotomous (dissimilarity calculated on a categorical variable). Il this latter expression $\bar{R}_1$ is the average of ranks associated to the values of $d_{ij,wG}$ for the subset of couples of units in which $d_{ijt} = 1$; similarly, $\bar{R}_0$ is the average of ranks associated to the $d_{ij,wG}$ for the subset where $d_{ijt} = 0$.

The major drawback of our proposal of searching weights $w_t$ with Brogden's estimator instead of point biserial correlation or by considering the rank's correlation (with biserial correlation coefficient in case of dissimilarity for categorical variables) is the non-negligible computational effort required for numerically searching via the genetic algorithm a solution to the problem (in both the cases the analytical approach does not work). The computational effort is expected to increase by increasing the number of variables (*p*) involved in the calculation of distance as well as the size of the final dissimilarity matrix (number of *m* couples of units).

## 4. Simulation study

This section presents results of some simulation studies related a *k*-NN classification application and to imputation of missing values in survey data using the nearest neighbor hotdeck approach. The simulation study involves both artificial and real survey data and is carried out in the R environment (R Core Team, 2023).

### *4.1. k-NN application*

In this case study we tackle a classification problem by applying the popular *k*-NN algorithm whereas dissimilarities are calculated with unweighted and weighted versions of the Gower's coefficient. The assessment uses artificial data generated using the facilities in the R package **clusterGeneration** (Qiu and Joe, 2020). In particular the sample consists of approximately 120 observations grouped in four clusters (clusters have a variable size randomly determined in a range from 20 to 40 units per cluster). For each cluster the values of two continuous variables are randomly drawn from a Gaussian distribution with 0 mean and a variance-covariance matrix that changes according to the cluster separation index: the closer to 1 the index is, the more separated the clusters are. In our experiment we consider a value of 0.3. Two categorical variables are added to the dataset by randomly changing the cluster membership of respectively a fraction of 0.2 and 0.4 of the observations; finally, a third a categorical "noisy" variable is added by randomly assigning the cluster membership to each unit.

The dataset is randomly split in a *training* subset (70% of whole sample) and a *test* subset (remaining 30% of units). The training data contribute to determine the weights that are then used in calculating the weighted Gower's dissimilarity. The weights are those that minimize the differences in the correlation between $d_{wG}$ and the dissimilarity calculated on each single variable ($d_t$). Four different alternative ways of measuring the correlation are considered:

- Pearson's correlation coefficient between $d_{wG}$ and each $d_t$ ("wPG"), as suggested by de Bello et al (2021);
- Pearson's correlation coefficient between $d_{wG}$ and $d_t$ when $d_t$ refers to a continuous variable and Brogden's estimator of point biserial correlation when $d_t$ refers to a categorical variable ("wpbG");
- Spearman's correlation coefficient between $d_{wG}$ and each $d_t$ ("wSG");

- Spearman's correlation coefficient between $d_{wG}$ and $d_t$ referred to continuous variables and rank biserial correlation when $d_t$ refers to a categorical variable ("wSbG");

The weights estimated on the training dataset are then applied to calculate the Gower's dissimilarity between the units in the test subset and those in the training subset and subsequently assign to each unit in the test dataset the most-voted cluster membership of the closest *k* nearest neighbors in the training set; the values 7, 9 and 11 are considered for *k*. Initially the noisy categorical variable is excluded, subsequently it is included.

Practically, 100 iterations of the procedure are done for each combination of the input parameters (presence/absence of the noisy variable; five different approaches for calculating the Gower's dissimilarity; three different values for the *k*). The evaluation of results is done by looking at the accuracy i.e. the capability of predicting the true cluster membership of each unit in the test dataset. Table 2 shows average results over 100 runs of the whole procedure. In it rows denoted with "unwG" indicates results achieved applying the standard unweighted Gower's dissimilarity coefficient.

Table 2 – Results of unweighted and weighted Gower's coefficient in *k*-NN classification

|  | Weights of categorical variables | | | Weights of continuous variables | | | Classification accuracy | | |
| --- | --- | --- | --- | --- | --- | --- | --- | --- | --- |
|  | P02 | P04 | noisy | V1 | V2 | noisy | k=7 | k=9 | k=11 |
| unwG | 0.2500 | 0.2500 |  | 0.2500 | 0.2500 |  | 0.8610 | 0.8666 | 0.8709 |
| wPG | 0.1479 | 0.2243 |  | 0.3148 | 0.3130 |  | 0.9005 | 0.8840 | 0.8783 |
| wPbG | 0.0791 | 0.1703 |  | 0.3760 | 0.3747 |  | 0.9476 | 0.9343 | 0.9309 |
| wSG | 0.1773 | 0.2454 |  | 0.2903 | 0.2870 |  | 0.8826 | 0.8738 | 0.8583 |
| wSbG | 0.0835 | 0.1780 |  | 0.3695 | 0.3691 |  | 0.9426 | 0.9260 | 0.9179 |
| unwG | 0.1667 | 0.1667 | 0.1667 | 0.1667 | 0.1667 | 0.1667 | 0.8067 | 0.7716 | 0.7482 |
| wPG | 0.0722 | 0.1109 | 0.1418 | 0.1615 | 0.1591 | 0.3545 | 0.7685 | 0.7457 | 0.7273 |
| wPbG | 0.0292 | 0.0775 | 0.1097 | 0.2031 | 0.2009 | 0.3795 | 0.8364 | 0.8081 | 0.7716 |
| wSG | 0.0770 | 0.1150 | 0.1421 | 0.1503 | 0.1496 | 0.3661 | 0.7625 | 0.7396 | 0.7107 |
| wSbG | 0.0367 | 0.0832 | 0.1136 | 0.1876 | 0.1871 | 0.3918 | 0.8077 | 0.7946 | 0.7603 |

Table 2 shows that the proposed weighing procedures determines an increase of weights assigned to continuous variables if compared to those of categorical variables. Introducing biserial correlation when one of the dissimilarities is calculated on a categorical variable produces a further increase of the weight associated to dissimilarities calculated on continuous variables. In absence of the noisy variable, the unequal weights contribute to a visible increase of the average classification accuracy; weights achieved when considering biserial correlation between the overall dissimilarity and the dissimilarity on categorical variables tend to perform better, while rank correlation produces slightly worst results. In general, increasing the value of *k* determines, in average, a slight reduction of classification accuracy[2].

---

[2] It is known that increasing the value of *k* makes boundaries between classes less distinct.

When the noisy variable is included in the calculation of dissimilarity as expected there is a decrease in average accuracy, more evident when increasing the value of *k* (if compared to results obtained excluding noisy variables from the calculation of the dissimilarity).

*4.2. Nearest neighbor donor imputation*

Nearest neighbor donor imputation (NNDI) is a very popular approach for imputing missing values in survey data; it falls in the class of *hotdeck* methods, i.e. methods that replace a missing value with a value observed on a unit sharing the same characteristics of that with the missing value (Andridge and Little, 2010). In practice, in NNDI the missing value is replaced with value observed on the closest observation in term of dissimilarity calculated on a set of auxiliary variables observed for both the units; the Gower's dissimilarity is the preferred choice when the selected variables include both categorical and continuous variables.

This section presents the result of a simulated study based on real survey data. In particular, it is considered the Survey of Household Income and Wealth in Italy, a sample survey carried out every two years by the Bank of Italy (2018). The disseminated data-files provide anonymized microdata[3] related to the about 8,000 Italian households (HHs), for the scope of this study we consider data related to year 2016 and in particular the subsample of 477 households consisting of just one person and living in the North-East of Italy. This subset is used for simulating nonresponse on the variable related to consumption expenditures and then fill-in the missing values using NNDI; in particular the following steps are considered:

1) The missing value on the variable related to consumption expenditures are randomly generated assuming probability of being missing equal to 0.5 for "employed" individuals and equal to 0.1 for the remaining ones; with the available data this device returns a fraction of missing values approximately equal to 0.25 and corresponds to a Missing At Random (MAR) mechanism (see. e.g. Little and Rubin, 2002), as the missingness depends on the employment status. Units with the missing values on the consumption expenditures form the subset of *recipients* while the remaining ones represent the *donors*.

2) Dissimilarities between recipients and donors are calculated considering the unweighted ("unwG") and the weighted Gower's dissimilarity, according to the various options presented in this article ("wPG", "wPbG", "wSG" and "WSbG"), using two variables: the net income (continuous) and the educational level (categorical with 8 levels); these variables are good predictors of the consumption expenditures (mainly the income). A larger set of variables is also formed by adding the marital status (categorical with 4 levels) and working status (categorical with 3 categories). These latter variables are not as good predictors of expenditures as the former ones but the working status determines the response probability.

---

[3] The anonymized microdata can be downloaded at:
https://www.bancaditalia.it/statistiche/tematiche/indagini-famiglie-imprese/bilanci-famiglie/distribuzione-microdati/index.html?com.dotmarketing.htmlpage.language=1

3) for each recipient it is imputed the value of consumption expenditures observed on the closest donor; when more donors are found at the minimum distance, for imputation purposes just one of them is picked at random.

All these steps are repeated 250 times and the assessment checks how the reconstructed dataset, formed by joining observed and imputed values (according to the various options), is able to reproduce the "true" total amount of consumption expenditures (estimated on the data before introducing the missing values in the step (1):

$$srB = \frac{1}{\hat{t}}\frac{1}{250}\sum_{b=1}^{250}[\tilde{t}_{obs\&imp,b} - \hat{t}]; \quad srRMSE = \frac{1}{\hat{t}}\sqrt{\frac{1}{250}\sum_{b=1}^{250}(\tilde{t}_{obs\&imp,b} - \hat{t})^2}$$

We also evaluate how imputed and observed data preserve the "true" distribution of consumption expenditures; this is roughly done by calculating the average differences between the quintiles estimated using imputed and observed values ($\hat{q}_{obs\&imp,bu}$) and the "true" quintiles $q_{bu}$ (estimated before deleting the simulated missing values); in particular, average absolute differences are considered:

$$sDQ = \frac{1}{250}\sum_{b=1}^{250}\left[\frac{1}{6}\sum_{u=1}^{6}|\hat{q}_{obs\&imp,bu} - q_{bu}|\right]$$

Table 3 – Results of unweighted and weighted Gower's coefficient in nearest neighbor donor imputation

|  | Weights in Gower's dissimilarity | | | | srB (x1000) | srRMSE (x1000) | sDQ |
|---|---|---|---|---|---|---|---|
|  | Net income | Edu. level | Marital status | Working status | | | |
| unwG | 0.5000 | 0.5000 |  |  | 3.9033 | 21.2641 | 1009.06 |
| wPG | 0.8451 | 0.1549 |  |  | 5.4617 | 22.0241 | 976.36 |
| wPbG | 0.9137 | 0.0863 |  |  | 6.4807 | 21.8843 | 964.55 |
| wSG | 0.5369 | 0.4631 |  |  | 3.9033 | 21.2641 | 1009.06 |
| WSbG | 0.9303 | 0.0697 |  |  | 6.5642 | 21.8813 | 964.13 |
| unwG | 0.2500 | 0.2500 | 0.2500 | 0.2500 | -6.9757 | 23.0608 | 1118.18 |
| wPG | 0.7125 | 0.1076 | 0.0923 | 0.0875 | -1.9708 | 22.5668 | 1106.00 |
| wPbG | 0.7919 | 0.0745 | 0.0719 | 0.0617 | 2.3091 | 22.2855 | 978.53 |
| wSG | 0.7767 | 0.0885 | 0.0662 | 0.0686 | 3.0678 | 23.0553 | 998.62 |
| wSbG | 0.8524 | 0.0523 | 0.0528 | 0.0426 | 1.7840 | 21.6097 | 956.43 |

Table 3 summarizes the results obtained over the 250 iterations of the simulated imputation with NND. As expected, the different options for searching automatically the weights of the variables in the Gower's dissimilarity, shift weighting towards the unique continuous variable – the net income – involved in the computation of the dissimilarity. In the "shorter" case, when dissimilarity is calculated using only two variables, unequal weighting seems to affect the estimation of the total amount of consumption expenditures (estimated with imputed and observed values) as there is a slight increase in both the simulation relative bias ("srB") as well as in simulation relative root MSE (srRMSE), with the exception of weights obtained by minimizing the differences in the Spearman's rank correlation "wSG". On the contrary, the introduction of weighting schemes in calculating the Gower's dissimilarity improves the estimation of quintiles ("sDQ"), and more in general the preservation of the overall distribution. The situation changes when the dissimilarity is calculated using four auxiliary variables, here the unequal weighting changes the sign of the relative bias (negative with the standard unweighted Gower's dissimilarity) and also its magnitude. The best result is obtained when using Spearmans' rank correlation or rank biserial correlation; the same outcome is obtained in terms of preservation of the whole distribution.

In this specific application, the introduction of a set of unequal variable weights in calculating the Gower's dissimilarity seems to perform better when considering a wider set of variables that does not necessarily include the best predictors of the variable to be imputed but includes the variables that determines the missingness mechanism.

## 5. Conclusions

Gower's proposal represents the most popular option for calculating dissimilarity between observations using a set of mixed type variables. This work investigates the problem of balancing the contribution of the various types of variables by introducing a weighting system that is determined automatically, following the idea of de Bello et al (2021), by minimizing the differences in the correlations between the weighted Gower dissimilarity and in turn the dissimilarity calculated on each single contributing variable. We introduce some variants to measure the correlation when dealing with a dissimilarity calculated on a categorical variable assuming only 0 and 1 values, as well as the possibility of measuring correlation between ranks. The simulation study shows that this proposal can be beneficial in different applications; our suggestion of considering biserial correlation when dealing with a dichotomous dissimilarity improves the results compared to the initial idea of using Pearson's linear correlation coefficient; this result holds true also when ranks are used in replacement of the origin single dissimilarities. However, in our study, shifting form linear correlation to rank correlation is beneficial only in the imputation problem where it is observed a general improvement of all the indicators considered for evaluating the performances of the different options for determining the weights. The price to pay when introducing biserial correlation or correlation between ranks is the computational effort required to find the optimal weights by means of the genetic algorithm; this effort increases with increasing number of couples of observations as well as the number of variables. The simulation study confirms that selecting the variables involved in the computation of the dissimilarity is a crucial decision that depends on the application; it is not the focus of this work, however results related to clustering and classification show very well that the inclusion of unnecessary "noisy" variables may cancel out the benefits of introducing a set of weights automatically determined.

As noted before, shifting to ranks when measuring the correlation between the dissimilarities in our small study is beneficial only in the case study related to the imputation of missing values. This result is interesting and relevant because nearest neighbor donor imputation is a very popular approach in sample surveys that often observe mixed type variable; for these reasons additional investigation is deserved.

More in general, results and proposals in this article may turn useful in other problems, not strictly related to the calculation of dissimilarity, where ranking of observations is the main objective, like, for instance, the case of the weighting schemes when deriving composite indicators.